\documentclass[aps,prb,reprint,groupedaddress]{revtex4-1}

\usepackage{graphicx} 
\usepackage{color}
\usepackage{float}

\begin{document}

\preprint{}

\title{High pressure neutron scattering of the magnetoelastic Ni-Cr Prussian blue analogue}


\author{D. M. Pajerowski}
\email{daniel@pajerowski.com}
\affiliation{NIST Center for Neutron Research, National Institute of Standards and Technology, Gaithersburg, Maryland 20899, USA}

\author{S. E. Conklin}
\affiliation{NIST Center for Neutron Research, National Institute of Standards and Technology, Gaithersburg, Maryland 20899, USA}

\author{J. Le\~ao}
\affiliation{NIST Center for Neutron Research, National Institute of Standards and Technology, Gaithersburg, Maryland 20899, USA}

\author{L. W. Harriger}
\affiliation{NIST Center for Neutron Research, National Institute of Standards and Technology, Gaithersburg, Maryland 20899, USA}

\author{D. Phelan}
\affiliation{NIST Center for Neutron Research, National Institute of Standards and Technology, Gaithersburg, Maryland 20899, USA}


\date{\today}

\begin{abstract}
This paper summarizes 0~GPa to 0.6~GPa neutron diffraction measurements of a nickel hexacyanochromate coordination polymer (NiCrPB) that has the face-centered cubic, Prussian blue structure.  Deuterated powders of NiCrPB contain $\approx$100~nm sided cubic particles.  The application of a large magnetic field shows the ambient pressure, saturated magnetic structure.  Pressures of less than 1~GPa have previously been shown to decrease the magnetic susceptibility by as much as half, and we find modifications to the nuclear crystal structure at these pressures that we quantify.  Bridging cyanide molecules isomerize their coordination direction under pressure to change the local ligand field and introduce inhomogeneities in the local (magnetic) anisotropy that act as pinning sites for magnetic domains, thereby reducing the low field magnetic susceptibility.
\end{abstract}


\maketitle



\section{Introduction}

A rich array of properties are displayed in coordination polymers (CPs).\cite{ka04}  In the realm of magnetism, many systems within this chemical motif have been selected by physicists to realize particular Hamiltonians that continue to increase understanding of fundamental issues.\cite{g04}  Here, we approach from the different vantage point of probing inherent properties of CPs that are expressed in functional systems.  Specifically, pressure dependent magnetism is an attractive property for transducers, and while it has been studied in metals and metallic alloys for centuries,\cite{j47} the more elaborately structured CPs are only recently being investigated.

The system of interest in this manuscript is the nickel hexacyanochromate (NiCrPB) CP that shows large changes in magnetic susceptibility with a modest application of pressure.\cite{zakklmmma07}  However, there is no reported, rigorous understanding of this response.  In 1992, the superlative, 90~K ferromagnetic ordering temperature ($T_{C}$) of NiCrPB fostered the continuing interest in magnetic CPs.\cite{gmcvp92}  Having spin-only ions and small inherent anisotropy, the magnetism of bulk NiCrPB could be proficiently modeled until 2007 when a factor of two reduction of the magnetization was seen with the application of 0.8~GPa in a field of 5~mT.\cite{zakklmmma07}  A resurgent interest in NiCrPB and its pressure dependent magnetism was sparked in 2010 when a new type of photomagnetic effect was observed in heterostructures of cobalt hexacyanoferrate (CoFePB) and NiCrPB.\cite{pagkmd10}  This year, subsequent reports definitively showed the hypothesized strain coupling between the CoFePB that has a photoinduced volume change and the neighboring layer of the magnetoelastic NiCrPB.\cite{prldd14}\cite{rqbalmd14}\cite{pmmclhlgrwbt14}  Somewhat puzzlingly, those works suggest that structural deformations relax within tens of nanometers of the surface while optimal heterostructures have NiCrPB layers that are hundreds of nanometers.

There are different, existing precedents in the literature that provide candidates to explain the magnetoelasticity of NiCrPB.  First, this effect was qualitatively explained as pressure induced tilting of metal-ion coordinated octahedra via linker buckling that subsequently couples to the local magnetic moments and reduces the magnetization component along the measuring axis via a site-by-site spin canting.\cite{zakklmmma07}  This type of structural modification is not surprising for a Prussian blue analogue (PBA) as pressure induced distortions are seen in CoFePB,\cite{bcbvmbj08} and recent X-ray absorption measurements provide further evidence for pressure induced deformations in nickel hexacyanoferrate (NiFePB).\cite{cliba13}  Second, it is possible that spin-canting is correlated on a longer length scale than site-by-site, and magnetostatic domain effects were invoked to explain anisotropy in thin films of NiCrPB.\cite{pgadgkhtm10}  Finally, NiCrPB may behave like iron hexacyanochromate (FeCrPB), which has a similarly large reduction in magnetization with applied pressure.\cite{cglrgmm05}  For FeCrPB, there is an isomerization of CN moeities that gives rise to a spin-transition from $d^6-$Fe$^{2+}$ (S~=~2) to $d^6-$Fe$^{2+}$ (S~=~0) as the Fe ligand field increases from N-coordination to C-coordination.

We have synthesized NiCrPB powders in heavy water and characterized their chemical make-up with X-ray photoelectron spectroscopy (XPS).  X-ray diffraction (XRD) and neutron powder diffraction (NPD) patterns were co-refined to give atomic coordinates within the network repeat unit and provide a scale factor for the magnetic scattering.  The pressure dependent NPD can directly interrogate the aforementioned hypotheses: as a function of pressure above $T_{C}$, it is sensitive to structural changes of metal ions and organic constituents, and below $T_{C}$ it can detect changes in the coherently averaged local moments.  These data taken together present a self-consistent model for the (pressure dependent) magnetizing process in NiCrPB.  One main feature we find is a change in the nuclear structure factor of NiCrPB with pressure that is best modeled as a structural isomerization of the CN molecules.  A second main feature we find is little change in the magnetic structure factor of NiCrPB with pressure that points to a domain reorganization model as the dominant modulator of the changes in magnetism. We support the validity of the proposed models with density functional theory (DFT) calculations and micromagnetic calculations.

\section{Methodology\cite{NISTdisclaimer}}
\subsection{Synthesis}
For synthesis, two continuously stirred and stoppered vials were connected via a peristaltic pump under N$_2$ atmosphere.  All chemicals were purchased from Sigma-Aldrich and used without further purification.  The source vial contained 0.194~g of NiCl$_2$ in 150~mL of heavy water, and the sink vial contained 0.335~g of KCl and 0.488~g of K$_3$Cr(CN)$_6$ in 300~mL of heavy water.  Solution transfers were performed at 3~mL/min.  Each sat for 12~hours before 1~hour of 4~krpm centrifugation at 23$^{\circ}$C (296 K).  The precipitate was dried under vacuum at 80$^{\circ}$C (353~K) to a fudge-like consistency to avoid contamination with non-isotopic water.  Four batches were combined for NPD studies.

\subsection{Instrumentation}
The XPS spectra were collected on a Kratos AXIS Ultra DLD equipped with a monochromatic, 140~W Al source (1486.6~eV) operating at 1~$\times$~10$^{−6}$~Pa (1~$\times$~10$^{−8}$~Torr).  Charge compensation used the neutralizing electron gun, aromatic C-1s levels were defined to be 284.7~eV, and the analyzer was 20~eV with a step size of 0.05~eV.  We used the relative sensitivity factors (RSFs) from Kratos for Cl and N, and derived our own for Ni using NiCl$_2$ and for K and Cr from K$_3$Cr(CN)$_6$.  The XRD was at room temperature (296~K) on a Rikagu Ultima III using a 1.6~kW Cu anode ($\lambda$~=~1.54~$\textrm{\AA}$).  Thermal NPD experiments were performed on BT-4 of the NIST Center for Neutron Research (NCNR) and cold neutron experiments were performed on the NG-5 Spin Polarized Inelastic Neutron Spectrometer (SPINS) of the NCNR.  Both machines used the (002) reflection of pyrolytic graphite (PG) as a monochromator and analyzer.  The SPINS experiment used 80’ collimators and the BT-4 experiment used 40’ collimators, with no collimation between analyzer and detector.  BT-4 was set at 14.26~meV ($\lambda$~=~2.395~$\textrm{\AA}$) with PG filters, and SPINS used 5.00~meV ($\lambda$~=~4.05~$\textrm{\AA}$) with a cold Be filter.  All NPD data were collected on constant monitor, at approximately 10 seconds per point for the large survey, 3 minutes per point for the high magnetic field data, and 12 minutes per point for the high pressure data.  Magnetic fields were applied with a superconducting magnet on BT-4 (0~T and 4~T), and an electromagnet on NG-5 (5~mT).  High pressure was achieved using a two-stage helium intensifier from Harwood Engineering with a 1.5~cm$^3$ aluminum alloy cell connected to the intensifier through a heated high-pressure capillary.  Pressure was adjusted at temperatures well above the helium melting curve and the capillary was heated during slow cooling of the cell to accommodate the contracting He gas, minimizing pressure loss.

\subsection{Calculations}
FULLPROF was used to refine the wide-angle diffraction patterns,\cite{r93} and the tabulated values for scattering lengths\cite{s92} and magnetic form factors\cite{cc74} were used when modeling the magnetic contribution.  High field data were analyzed using the same rubric as for CoFePB.\cite{pgkadcnttm12}  Spin-polarized DFT calculations with LDA-functionals used the GPAW\cite{mhk05}\cite{erm10} and ASE\cite{bk02} codes. The real-space grid had a nominal spacing of 0.15~$\textrm{\AA}$ on a (64, 64, 64) grid for the crystal calculations and 0.2~$\textrm{\AA}$ on a (80, 80, 80) grid for the molecular calculations.  The unit cell (also used for crystal calculations) contains four times as many atoms as the chemical formula. The criterion of convergence for DFT calculations was applied without any symmetry constraints to be $\leq$10$^{–5}$~eV/electron in energy for crystal calculations and $\leq$10$^{–6}$~eV/electron for molecular calculations, $\leq$4~$\times$~10$^{–8}$ eV$^2$ for integrated eigenstate change, and $\leq$0.01~eV/$\textrm{\AA}$ for residual interatomic forces. Magnetic moments were initially set to the expected single-ion values and relaxed during optimization.  Zero temperature micromagnetic simulations used the three dimensional implementation within the OOMMF package\cite{dp99} with 100~nm sided cubic particles divided into 1000~nm$^3$ micromagnetic volumes.   To choose magnetic fields in powder averaged micromagnetic simulations, the cubic particle was reduced to the highest symmetry wedge and the distance between nine points within that area was maximized numerically.

\section{Nuclear Structure}
\begin{figure}[t!]
	\includegraphics{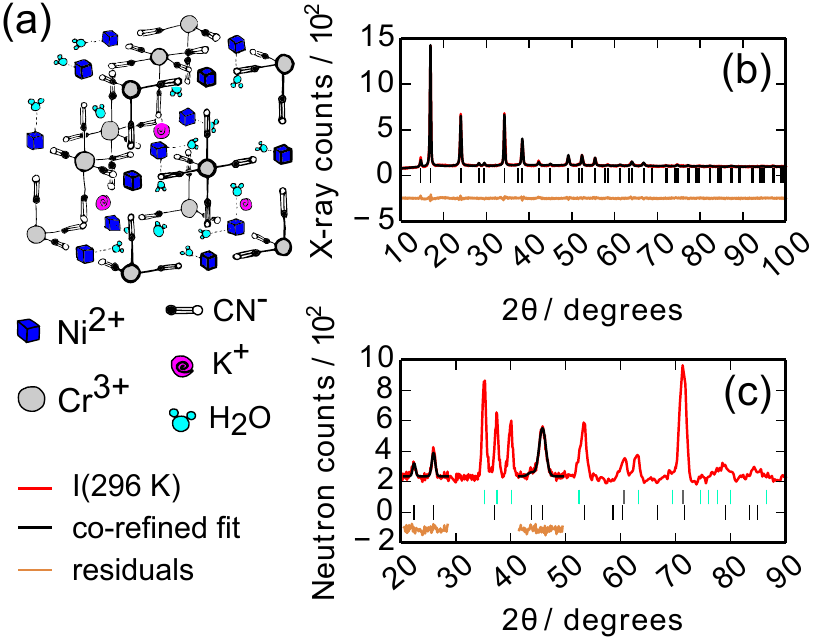}
	\caption{(color online) Nuclear structure of NiCrPB.  (a) The NiCrPB unit cell is quantified with (b) XRD and (c) NPD, both at 296 K.  Black ticks indicate NiCrPB reflections, aqua ticks indicate D$_2$O reflections, and gray ticks show aluminum reflections.}
	\label{fig:NiCr_figure1}
\end{figure}

We begin the analysis with chemical characterization, XRD, and thermal NPD.  From XPS, we arrive at a chemical formula of K$_{0.42}$Ni[Cr(CN)$_6$]$_{0.88}$$\cdot$xD$_2$O, which has a negative net charge of 0.22 electrons and we do not use XPS to analyze the water content due to the vacuum atmosphere.  Starting by substituting nickel and chromium into the face-centered cubic (s.g. 225 $Fm\bar{3}m$) Prussian blue structure,\cite{bspa77} Fig.~\ref{fig:NiCr_figure1}(a), the 296~K XRD and NPD were co-refined, Fig.'s~\ref{fig:NiCr_figure1}(b)-(c), to give a slightly different chemical formula of K$_{0.25}$Ni[Cr(CN)$_6$]$_{0.75}$(D$_2$O)$_{0.25}$$\cdot$2.1D$_2$O that has a net charge of 0.0.  To reduce the number of fitting parameters, the CN and D$_2$O intermolecular distances are constrained \cite{bs11} and only one Debye-Waller factor is used for the crystal (B~=~3.6~$\textrm{\AA}^2$) with an additional Debye-Waller factor (B'~=~10~$\textrm{\AA}^2$) for the disordered interstitial heavy water molecules.   Only those peaks in NPD that are separate from both the aluminum holder signal and heavy water signal were co-refined.  The diffracted beam line-widths are consistent with $\approx$100~nm crystallites from a cubic particle Scherrer equation analysis,\cite{am76} which is in accord with previous TEM reports from similar synthesis.\cite{dkgpgkmd11}  Refined positions of atoms within the unit cell are reported in Table~\ref{table1} and the derived scale factor for the NPD is then used for magnetic scattering.

\begin{table} []
\caption{Atomic coordinates and occupancies for NiCrPB at T~=~296~K.  Space group $F m \bar{3} m$ (no. 225), a = 10.484 $\textrm{\AA}.$\\\label{table1}}
\begin{ruledtabular}
\begin{tabular}{ c c c c c c}
\ Atom &	Position	&n	&x&	y&	z \\
\hline
Ni & 4a & 1.00 & 0.50 & 0.50 & 0.50\\
Cr & 4b & 0.75 & 0.00 & 0.00 & 0.00\\
C & 24e & 0.75 & 0.19 & 0.00 & 0.00\\
N & 24e & 0.75 & 0.30 & 0.00 & 0.00\\
K & 8c & 0.20 & 0.25 & 0.25 & 0.25\\
(D$_2$O) & 24e & 0.25 & 0.31 & 0.00 & 0.00\\
(D$_2$O) & 32f & 0.33 & 0.31 & 0.31 & 0.31\\
\end{tabular}
\end{ruledtabular}

\end{table}

\section{High Field Magnetic Structure}
\begin{figure}[t!]
	\includegraphics{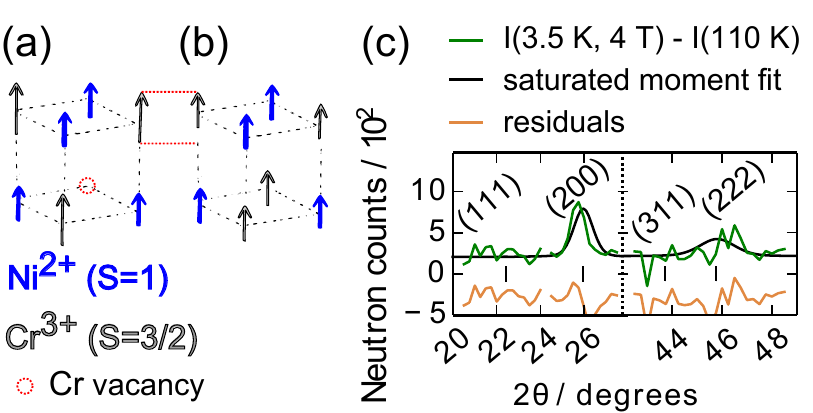}
	\caption{(color online) High field magnetic structure of NiCrPB.  (a) The high field magnetic structure coincidentally balances chemical occupation with spin magnitude to give (b) a magnetic unit cell that is approximately half of the chemical cell, as shown by the nature of (c) the additional scattering at 3.5~K and 4~T after subtracting Debye-Waller corrected data at 100~K.  The ticks indicate NiCrPB reflections.}
	\label{fig:NiCr_figure2}
\end{figure}

During the same thermal NPD experiment, the temperature and applied magnetic field were changed to probe the high field magnetic structure.  By comparing relative NPD intensities at 296~K and 100~K and using the Debye model,\cite{am76} we find a Debye temperature of 270~K that is used to remove thermal effects when comparing patterns above and below TC (the temperature decrease from 100~K to base temperature reduces thermal displacements, $\left\langle u\right\rangle^2$, by ≈1.75 times). At low temperature and high magnetic field, 3.5~K and 4~T in this measurement, the magnetization in NiCrPB saturates as a ferromagnet \cite{gmcvp92} and can be modeled without consideration of domains and magnetic nanostructure.  According to magnetometry, magnetic resonance, and ligand field analysis, the magnetic terms are spin-only Ni$^{2+}$ ($d^8$, $^3A_2$, S~=~1) and Cr$^{3+}$ ($d^3$, $^4A_2$, S~=~3/2) and the g-factor is nearly 2.\cite{gmcvp92}\cite{pgadgkhtm10}  Interestingly, the average Cr$^{3+}$ moment as approximately the same magnitude as the average Ni$^{2+}$ moment because the larger magnetic moment of Cr$^{3+}$ is almost exactly offset by Cr(CN)$_6$ vacancies, Fig.~\ref{fig:NiCr_figure2}(a).  Thus, the neutrons measure a magnetic unit cell that is approximately simple cubic, Fig.~\ref{fig:NiCr_figure2}(b).  This happenstance is manifest by the fact that only second order reflections are observed when the nuclear background as been subtracted from the low temperature, high field data (T~=~3.5~K, $\mu_0$H = 4 T) as shown in Fig.~\ref{fig:NiCr_figure2}(c), which indicates that the magnetic cell is metrically half the size of the nuclear cell.  As a result, the NPD is consistent with the bulk measurements and NiCrPB can be considered a simple cubic ferromagnet of S~=~1 for diffraction purposes.

\section{High Pressure Neutron Diffraction}
\begin{figure}[t!]
	\includegraphics{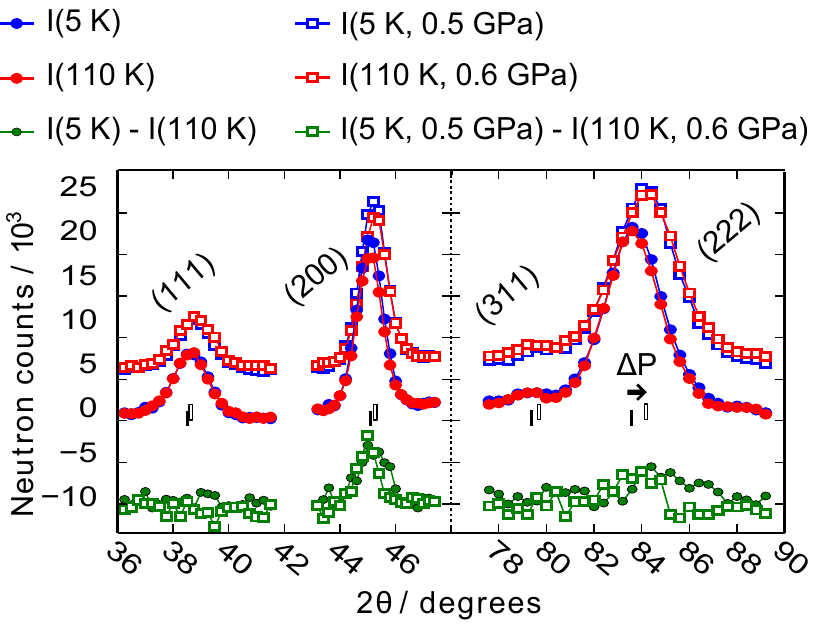}
	\caption{(color online) High pressure NPD of NiCrPB.  High pressure data are offset by $5~\times~10^3$ counts, and subtraction data are offset by $-10^4$ counts and scaled by a factor of three.  Black tick marks are peak positions at ambient pressure, open tick marks are peak positions at high pressure.}
	\label{fig:NiCr_figure3}
\end{figure}

We investigated the response of NiCrPB to high pressure by performing neutron diffraction using the lower background, cold triple-axis.  Four conditions were measured in $\mu_0$H = 5 mT: ($P_0$-cold) below $T_C$ at 5~K and at $\approx$100~kPa, ($P_0$-hot) above $T_C$ at 110~K and at $\approx$100~kPa, ($P_{0.5}$-cold) below $T_C$ at 5~K and at 0.50~GPa, and ($P_{0.6}$-hot) above $T_C$ at 110~K and at 0.60~GPa.  The $P_{0.6}$-hot and $P_{0.5}$-cold pressures were chosen due to practical restrictions.  These data are plotted together in Fig.~\ref{fig:NiCr_figure3}, along with scaled subtractions of [$P_0$-hot - $P_0$-cold] and [$P_{0.6}$-hot - $P_{0.5}$-cold] that show the additional magnetic scattering has a pattern superficially similar to the high field case.  In the following, we first quantify the pressure dependent structural changes and then the pressure dependent changes in magnetic scattering are detailed.

\subsection{Structural Scattering}
First, the positions of the peaks are sensitive to the size of the unit cell, which gives one measure of how NiCrPB responds to strain.  There is very little thermal contraction, but pressure has a drastic effect on the peak positions that can be used to derive the bulk modulus, $K~=~-VdV/dP$.  The 110~K phase shows a change from 10.477~$\textrm{\AA}$ at ambient pressure to 10.410~$\textrm{\AA}$ at high pressure, such that $K~=~31.43$~GPa.  Similarly, at 5~K the application of pressure causes a contraction from 10.468~$\textrm{\AA}$ to 10.413~$\textrm{\AA}$ that gives a nearly identical value of $K~=~31.94$~GPa.  The ansatz of linear volume contraction with pressure is supported by the similar $K$ values for 0.5~GPa and 0.6~GPa.  As a frame of reference, these modulus values are slightly less than CoFePB (43~GPa),\cite{bcbvmbj08} near those for silica glass (35~GPa to 55~GPa),\cite{sa91} and considerably less than the 170~GPa of elemental iron.\cite{l03}

\begin{figure}[t!]
	\includegraphics{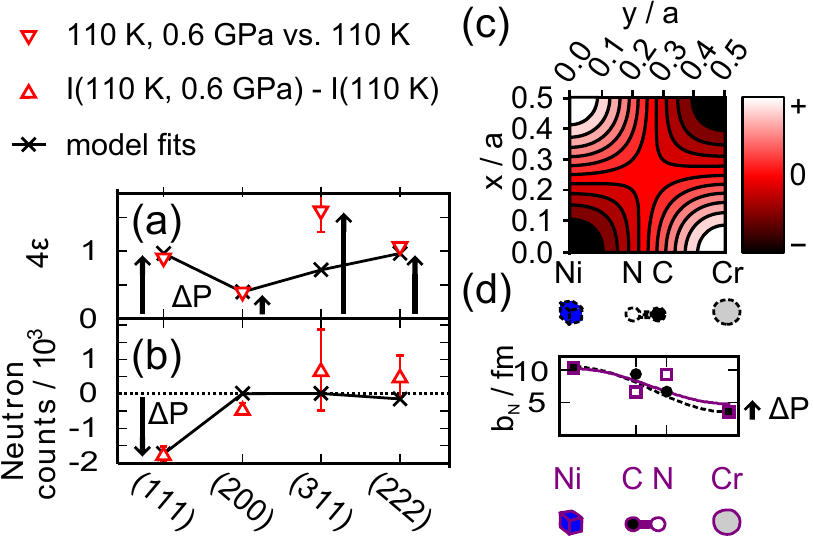}
	\caption{(color online) Pressure induced structural changes in NiCrPB.  (a) The strain, 4$\varepsilon$, increases with pressure and the fit to an anisotropic cubic strain model is shown.  (b) The intensity of the (111) peak changes the most, and the 20\% isomerized $\mu$-CN model fit is shown.  For (b) and (c), uncertainty bars come from least squares fits to the peaks.  (c)  A Fourier transform of the change in scattering length density with pressure is in the unit cell of NiCrPB at the z = 0, x-y plane that contains metal ions and CN molecules with fractional coordinates as per Table~\ref{table1}.  (d) The conventional cyanide linkage, Ni-N-C-Cr, and an isomerized cyanide linkage, Ni-C-N-Cr, are illustrated with horizontal positions aligned to the above Fourier difference map.  The bound neutron scattering length, $b_N$, for the atoms is shown along with a fit to a cosine function in the z=0, y=0, x-$b_N$ plane that illustrates the decrease in (111) wave amplitude with isomerization, although quantitative modeling allows for interference of all atoms within the cell.}
	\label{fig:NiCr_figure4}
\end{figure}

Second, the widths of the peaks have information about particle size and strain, and these data show pressure induced anisotropic strain broadening.  By taking the geometrical difference of widths between low pressure ($\beta_0$) and high pressure ($\beta_P$) for the non-magnetic high temperature phase, a width associated with pressure induced strain broadening may be scrutinized ($\beta_{strain}~=~(\beta_P^2-\beta_0^2)^{1/2})$.  Strain, $\varepsilon$, has an effect on line-width ($\beta_{strain}~=~4\varepsilon tan\theta$) that may be expressed as variances of lattice spacings, $d_{hkl}$, such that 4$\varepsilon$ = $d_{hkl}^{2}(\sigma^2(d_{hkl}^{-2}))^{1/2}$.\cite{s99}   For a cubic system, these variances have two parameters and we find $S_{400}$ = 0.16~$\textrm{\AA}^{-4}$ and $S_{220}$ = 2.70~$\textrm{\AA}^{-4}$ reproduce the anisotropic experimental behavior of NiCrPB, Fig.~\ref{fig:NiCr_figure4}(a).  A similar anisotropic response to pressure was seen in CoFePB.\cite{bcbvmbj08}

Third, the intensities of the peaks provide information about the fractional coordinates of atoms within the unit cell and their positional distributions.  We consider the high temperature data to avoid magnetic scattering.  Fitting the peaks and subtracting the areas of $P_0$-hot from $P_{0.6}$-hot shows how the intensities are affected by pressure, Fig.~\ref{fig:NiCr_figure4}(b), where the intensity of the low angle (111) peak is changed the most.  Since the overall Debye-Waller factor does not appreciably change under pressure, there must be a correlated, strained contraction of the lattice with a coherent change in the structure factor.  Taking the Fourier transform of the subtracted intensities from Fig.~\ref{fig:NiCr_figure4}(b) provides a real-space visualization of how pressure changes the neutron scattering length density (SLD) in NiCrPB.  We show a two dimensional cut of this mapping onto the real-space crystallographic cell in the plane that contains metal ions and briding CN molecules, Fig.~\ref{fig:NiCr_figure4}(c), which shows increased SLD near Cr sites compared to Ni sites.  Taking a brief aside, understanding of the pressure induced change is helped by conceptualizing the (111) peak, which is due to SLD oscillations that have peaks and valleys at the 4a and 4b Wyckoff positions.  In the XRD of NiCrPB, Fig.~\ref{fig:NiCr_figure1} (b), the (111) is relatively weak because the dominant scatterers are Ni (28 electrons, 4a site) and Cr (24 electrons, 4b site), which have similar X-ray bound scattering lengths.  Conversely, in the NPD, e.g. Fig.~\ref{fig:NiCr_figure1} (c), the (111) relative intensity is stronger, due to the fact that the chain of atoms along the (111) oscillation ridge is mostly Ni-N-C-Cr, which have greater contrast and are arranged in descending order with respect to neutron scattering length to provide the appropriate, (111)-like, oscillatory behavior.  Analogously, the (200) arises due to SLD oscillations that have peaks at 4a and 4b sites and valleys at 8c sites, such that XRD has the stronger relative (200).  So, as a function of pressure, there must be an antisymmetric change in the SLD that increases in the vicinity of the Cr sites while simultaneously decreasing by a similar amount at the Ni site; if there was only an increase in SLD at the Cr site (such as if pressure inducing He gas were to load a vacancy site) the (200) would increase as the (111) decreases, which is not observed.  Therefore, something is happening on the 24e sites, which include C, N, and the coordinated heavy water molecules, Table~\ref{table1}, that have bound scattering lengths of 6.65~fm, 9.36~fm, and 19.14~fm, respectively.\cite{s92}  One change that can explain these data is an isomerization of the cyanide linker from carbon bonding to Cr to carbon bonding to Ni, Fig.~\ref{fig:NiCr_figure4}(d).  Chains of Ni-C-N-Cr have less of a (111) contribution than Ni-N-C-Cr because the SLD is less like a (111) plane wave.  Indeed, a change of 20\% of the CN to its structural isomer in the sample can reproduce the observed intensity change, Fig.~\ref{fig:NiCr_figure4}(b).  A choreographed shift of the coordinated water and the cyanide linker can also reproduce an antisymmetric change in SLD between the metal ions, but we were unable to find a quantitative model.

\subsection{Magnetic Scattering}
\begin{figure}[t!]
	\includegraphics{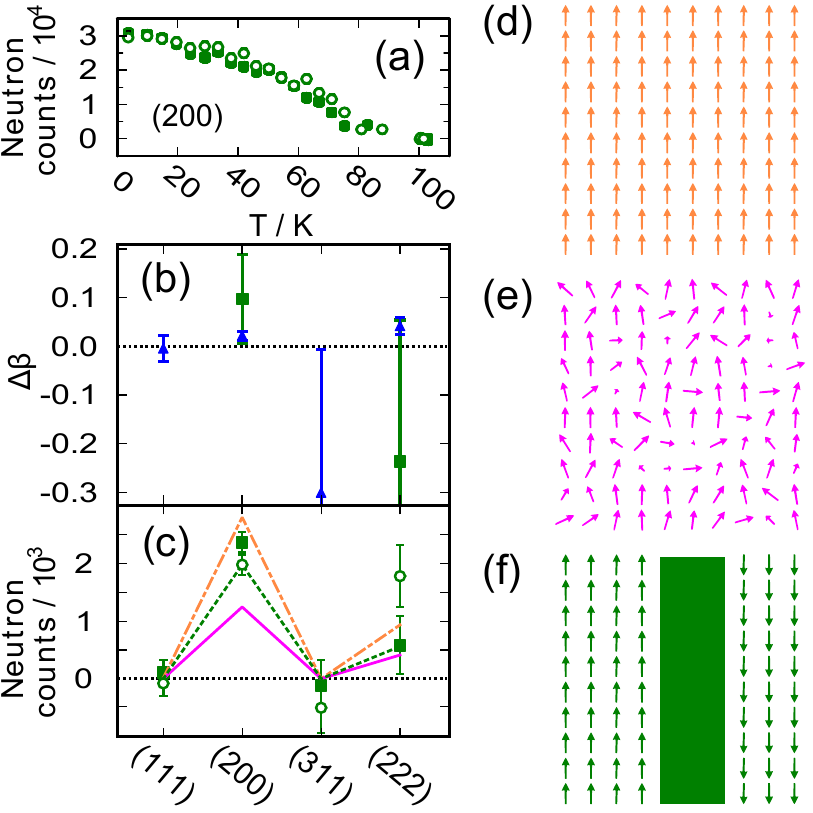}
	\caption{(color online) Pressure induced magnetization changes in NiCrPB. (a) The temperature dependence of the order parameter taken from the (200) reflection, with closed symbols for ambient pressure and open symbols for high pressure.  (b) For ambient pressure, the increase in peak width upon cooling is shown in blue triangles, and the increase of peak width specifically for the magnetic scattering is shown in green squares.  (c) The peak area due to magnetic scattering is shown in closed green squares for ambient pressure and open green circles for high pressure.  Uncertainty bars come from least squares fits to the peaks.  The dot-dashed burnt sienna line is assuming monodomain particles with spins along cubic axes (as in (d)~), the solid magenta line is assuming uncorrelated reduction in magnetization (as in (e)~), and the dashed green line is for a magnetic domain model (as in (f)~).  Magnetization illustrations represent two dimensional cuts through the center of a particle.  For (b) and (c), uncertainty bars come from least squares fits to the peaks.}
	\label{fig:NiCr_figure5}
\end{figure}

The additional scattering present below $T_C$ is confirmed to be magnetic by the temperature dependence of the (200) reflection, Fig.~\ref{fig:NiCr_figure5}(a), which indicates that $T_C~\approx$~75~K, consistent with that expected.\cite{zakklmmma07}\cite{gmcvp92}  The magnetic Bragg reflections do not strongly change with pressure in the way that the net magnetization does.\cite{zakklmmma07}  We note that an additional set of measurements of transmitted beam depolarization (not shown) that is sensitive to ferromagnetic ordering were performed that revealed a similar $T_C$  and a stronger pressure dependence.  However, even though the depolarization measurements suggest a reduction of moment along the field axis with pressure, unresolved questions about the neutron coherence length within the polarization setup precluded a quantitative analysis.

The widths of the diffraction peaks are used to estimate the size of the coherent magnetic domains.  Unfortunately, only the ambient pressure measurements can be compared, as the high pressure phases can also have changes due to their different relative pressure induced strain (i.e. the high pressure data at low temperature are 0.5~GPa and at high temperature they are 0.6~GPa).  For the ambient pressure data, the purely structural peak widths do not change with temperature, while the magnetic peaks indeed broaden.  This broadening is more pronounced when the structural scattering is subtracted out.  If changes in peak width, $\Delta\beta$, are only due to changes in magnetic versus structural domain size, $D$, for a cubic particle from differentiating the Scherrer equation\cite{am76} $\Delta D = -1.06 \times \Delta\beta cos(\theta) D^2/\lambda$, where $\theta$ is the scattering angle and $\lambda$ is the wavelength.  Fig.~\ref{fig:NiCr_figure5}(b) shows the broadening of peak widths and the magnetic contribution to this broadening.  For the (311) peak, there is a higher level uncertainty due to the proximity of the stronger (222), and the subtracted (222) has a higher level of uncertainty due to the much stronger relative contribution of nuclear scattering compared to magnetic scattering at this position.   Taking the width of the (200) peak that is well determined, if the structural domains are 100~nm, the magnetic domains would be 68~nm~$\pm$~29~nm.

The intensities of the magnetic Bragg reflections depend upon the magnitudes of the magnetic moments along with their correlations.  Since the nuclear structure has been determined, a scale factor can be derived in which the intensity of the observed magnetic scattering can be compared to different models.  Taking the high field structure and assuming a monodomain structure oriented along crystallographic axes would yield intensities of the four Bragg reflections probed as shown in Fig.~\ref{fig:NiCr_figure5}(c) (shown in burnt sienna).  Although the fit is of a reasonable quality and reproduces the relative moments of Ni$^{2+}$ and Cr$^{3+}$, the predicted intensity of the (200) reflection is larger than that observed under both low and high pressure.  Therefore, there is a loss of coherent magnetization perpendicular to the scattered wave vector.  If spins were to cant with pressure on a site-by-site basis, reducing the overall coherent moment by 30\% as in magnetometry \cite{zakklmmma07}(but assuming a tight distribution so the magnetic Debye Waller is unaffected), Fig.~\ref{fig:NiCr_figure5}(e) (shown in magneta), there would be significantly less scattering than observed such that we rule out this model as a correction to the monodomain model.  However, if a domain structure is introduced, coherent scattering will reduce depending upon the domain wall thickness, Fig.~\ref{fig:NiCr_figure5}(f) (shown in green), and a domain model can yield an arbitrary value for the net magnetization with a similar NPD pattern, until saturation when the domain wall annihilates.  A domain structure would also increase the magnetic peak width as the size of coherent magnetic regions is reduced, as we observed in the preceding paragraph.

\section{Discussion}
These NiCrPB NPD experiments confirm the nuclear structure and spin states, show clear changes in the nuclear structure with pressure, demonstrate differences in magnetization between high and low field, and manifest modifications to magnetization between ambient and high pressure.  The pressure induced structural changes affect the ligand field in NiCrPB, but without obvious additional buckling of the linkages (rotation of octahedra) such that there is no modification of the ferromagnetic superexchange, $J$, as when $>$~1~GPa.\cite{pltm13}  A stable form of CoFePB showed departure from the native cubic symmetry at $\approx$2~GPa.\cite{bcbvmbj08}  X-ray absorption studies of NiFePB suggest buckling of linkages at 1.6~GPa; strikingly there is no magnetoelastic effect in NiFePB up to 1 GPa (at 100 mT), due to an absence of CN isomerism for hexacyanoferrate species that may occur in the less stable hexacyanochromate (where stability is in comparison with cyanonickelate species).\cite{cliba13}  There emerges two regimes for pressure response in these cubic PBAs that has a cross-over at 1~GPa to 2~GPa.   So, while pressure in the first regime induces ligand field changes in the hexacyanochromate based FeCrPB that dramatically alter the Fe$^{2+}$ ground term,\cite{cglrgmm05} similar ligand field changes for Ni$^{2+}$ and Cr$^{3+}$ in NiCrPB do not affect the ground spin state but rather the orbitally non-degenerate excited states, which modify the g-factor and the magnetocrystalline anisotropy of NiCrPB.\cite{g61}  The introduction of local magnetocrystalline anisotropy also explains the loss of magnetic resonance intensity in photomagnetic strain coupled NiCrPB heterostructures near the g~=~2 position.\cite{k13} There is a randomness to ligand fields (thus anisotropy) in cubic PBAs due to chemical disorder, Fig.~\ref{fig:NiCr_figure1}(a).  In CoFePB, random local anisotropy dominates the magnetization determination, leading to a short magnetic correlation length ($L_C$).\cite{pgkadcnttm12}  Compared to CoFePB, NiCrPB has five times greater $J$ and five times less coercive field, $H_C$, increasing $L_C$ by orders of magnitude because neighbor spin misalignment becomes more costly than anisotropy axis misalignment.  Therefore, the reason that the NiCrPB magnetic NPD signal does not have a large intensity change while the magnetometry signal does is due to pressure causing spin rotations that are correlated at length scales on the order of the particle size; the observed small change in intensity and peak broadening of the magnetic phase is then due to the presence of a magnetic domain structure.  To probe plausibility of the proposed picture, we check the energetic viability of CN isomerism and look for local anisotropy with DFT, and assess the effect of such anisotropy on a particle-wide extent with micromagnetics.

\subsection{Density Functional Theory Calculations}
\begin{figure}
	\includegraphics{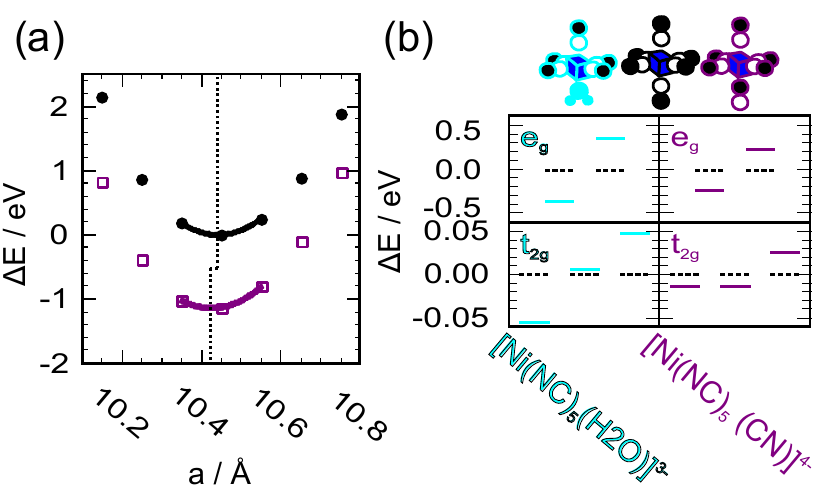}
	\caption{(color online) DFT calculations. (a) The energy as a function of lattice constant for “idealized” KNi[Cr(CN)6], solid black circles, and “isomerized” KNi[Cr(NC)6], open purple squares, where lines are parabolic fits used to extract the minimum energies, equilibrium lattice constants, and bulk moduli.  (b) The energy splittings of the majority spin electrons are shown relative to the center of mass of each symmetry block for selected molecules.  The minority spin levels are ≈1.0 eV higher in energy such that these energy shifts do not affect spin pairing.}
	\label{fig:NiCr_figure6}
\end{figure}

For DFT crystal calculations, “idealized” KNi[Cr(CN)$_6$] and “isomerized” KNi[Cr(NC)$_6$] are used because the defect lattice has structural glassiness that is problematic for DFT structural optimization.  To determine the required additional electron repulsion for the d-electrons (parameterized as Racah's “A” or Hubbard's “U” parameters),\cite{mhk05}\cite{g61} we use one parameter for both Ni and Cr and optimize the fractional coordinates while varying the lattice parameter of KNi[Cr(CN)$_6$].  This approach yields a nearly linear dependence of a~=~10.205~$\textrm{\AA}$~+(0.0315~$\times$~$U$) \textrm{\AA}/eV, so we choose $U~=~7$~eV.  The energy as a function of lattice constant shows equilibrium values of 10.442~$\textrm{\AA}$ and 10.426~$\textrm{\AA}$ for “idealized” and “isomerized” models, respectively, with the KNi[Cr(NC)$_6$] minimum lower by 1.1~eV per unit cell or 0.3~eV per chemical formula unit (Fig.~\ref{fig:NiCr_figure6}(a) ).  The different CN coordinations are nearly degenerate and energies not included in our simulation may stabilize one.  Notably, the “isomerized” state has a subtly decreased lattice constant.  The bulk moduli, $K_{DFT}~=~(1/9$a$)(d^2E/d$a$^2)$, are 73.62~GPa and 75.02~GPa for the “idealized” and “isomerized” crystals; scaling $K_{DFT}$ by the number of CN linkers in K$_{0.25}$Ni[Cr(CN)$_6$]$_{0.75}$ improves agreement with NPD but still overestimates experiment by 50\%.  Throughout compression and elongation over this range, the linkers remain straight, although at less than $\approx$10~$\textrm{\AA}$ the CN molecules buckle and the energy surface of cell volume becomes less defined (the DFT remains well-behaved while structural optimization does not).

The defects are investigated by discrete, charged molecular DFT calculations.  The local environments of metal ions in NiCrPB are estimated by [Cr(CN)$_6$]$^{3-}$, [Ni(NC)$_6$]$^{4-}$, and [Ni(NC)$_5$(H$_2$O)]$^{3-}$, for which DFT gives octahedral splitting, $\Delta _{oct}$, values of 4.7946~eV, 1.5634~eV, and 1.2877~eV, respectively.  The [Ni(NC)$_5$(H$_2$O)]$^{3-}$ species is present at the surface and near chromium vacancies, and we impose symmetry to avoid the strong attraction between the H and N atoms that is present in the molecular form.  Under pressure in the dilute isomerization limit that we are near at ≈0.5 GPa, CN may flip to give rise to [Ni(NC)$_5$(NC)]$^{4-}$, a-[Ni(NC)$_4$(NC)(H$_2$O)]$^{3-}$, and b-[Ni(NC)$_4$(NC)(H$_2$O)]$^{3-}$, where a- and b- denote parallel or perpendicular flips with respect to the water coordination axis.  In tractable terms, when an axial ligand changes in a transition metal coordination sphere, the different $\pi$-bonding of the alien ligand can shift $d_{xz}$ and $d_{yz}$ energies and the energy of $d_z^2$ can shift due to $\sigma$-bonding changes.  In the spectrochemical series, CN$^-$ is a strong $\pi$-acceptor, NC$^-$ a very weak $\pi$-acceptor, and H$_2$O a weak $\pi$-donor, while the $\sigma$-bonding presumably tracks the Lewis basicity.\cite{fh00}  The effect of such ligand field distortions on the d-electron energies are quantified here with DFT, and select molecules are shown in Fig.~\ref{fig:NiCr_figure6}(b).  Taking the DFT wavefunctions for these distorted geometries, the spin-orbit interaction ($\lambda_{SO}\vec{L}\cdot\vec{S}⃗$) then breaks the spin degeneracy of the ground terms via second-order non-degenerate perturbation theory, which is typically captured as $\propto$D$S_z^2$ (D is the anisotropy energy and the $z$ is the unique axis).\cite{g61}  This scheme mixes in the first excited state with non-zero matrix elements, which is an anisotropic, orbital triplet for both Ni$^{2+}$/$^3A_2$ and Cr$^{3+}$/$^4A_2$, to give an energy shift that depends upon the strength of the distortion and $\propto\lambda_{SO}^2$/$\Delta _{oct}$ and a g-factor shift by $\approx$8$\lambda_{SO}^2$/$\Delta _{oct}$.  Because free-ion values of $\lambda_{SO}$ for Cr$^{3+}$ and Ni$^{2+}$ are 34~meV and 80~meV, respectively, (although covalency can reduce the orbital moment) and Ni$^{2+}$ has the weaker ligand field, the anisotropy in NiCrPB is dominated by nickel ions.  For b-[Ni(NC)$_4$(NC)(H$_2$O)]$^{3-}$ the symmetry is lower than axial and the perturbation theory is more complicated, but it is clear that the symmetry axis is no longer a crystallographic axis.  For [Ni(NC)$_5$(H$_2$O)]$^{3-}$, we estimate D$_{CALC}$~=~0.247~meV, and for [Ni(NC)$_5$(NC)]$^{4-}$, D'$_{CALC}$~=~-0.092~meV.  Although on a single particle level, these energies are perturbative and will not be relevant for powder averaged paramagnetic susceptibilities, in the many body state they become significant.  Furthermore, these shifts are smaller than the spin-pairing energy such that no spin transition occurs.  However, if ligand field distortions are extreme enough, S~=~0 species may be stable, like in square-planar Ni$^{2+}$,\cite{fh00} and reduce the ordering temperature.

\begin{table*} 
\caption{Micromagnetic simulation parameters of NiCrPB with bulk Fe parameters for comparison.  These parameters are saturation magnetization, $M_{SAT}$, exchange stiffness, A, anisotropy, $K_1$, domain wall width, $\delta$, domain wall energy, $\gamma$, exchange length, $L_{EX}$, magnetic hardness, $\kappa$, and critical single-domain radius, $R_{SD}$.  The approximation symbols are included for NiCrPB parameters that depend upon $K_1$; moreover such parameters become less quantitative for highly inhomogeneous $K_1$-distributions.\\\label{table2}}
\begin{ruledtabular}
\begin{tabular}{ c c c c c c c c c}
 & $M_{SAT}$ & A & $K_1$ & $\delta$ & $\gamma$ & $L_{EX}$ & $\kappa$ & $R_{SD}$\\
 & A/m & J/m & J/m$^3$ & nm & J/m$^2$ & nm &  & nm\\
\hline
NiCrPB & 1.4 $\times$ 10$^5$ & 5.1 $\times$ 10$^{-13}$ & $\approx$ 1 $\times$ 10$^4$ & $\approx$ 7 & $\approx$ 7 $\times$ 10$^{-5}$ & 4.6 & $\approx$ 0.6 & $\approx$ 100\\
Fe\cite{s95} & 1.8 $\times$ 10$^6$ & 1.0 $\times$ 10$^{-11}$ & 4.8 $\times$ 10$^4$ & 40 & 2.6 $\times$ 10$^{-3}$ & 1.6 & 0.12 & 7\\
\end{tabular}
\end{ruledtabular}

\end{table*}

\subsection{Micromagnetic Calculations}
While NPD points towards the long $L_C$ and presence of domains in NiCrPB, additional magnetization studies will be necessary to uncover details of the domain structure (whether through imaging, single crystal work, studying a series of NiCrPB systems, or some other technique).  In the meantime, some insight into the magnetic nanostructure can still be gained from micromagnetics.  Micromagnetic simulations of NiCrPB include implementations of Zeeman energy, superexchange energy, magnetocrystalline anisotropy, and magnetostatic energy.\cite{dp99}  Normalizing to the measured unit cell volume, superexchange stiffness (A) is derived from the mean-field expression for $T_C$, and saturation magnetization ($M_{SAT}$) from the high field NPD.  Magnetocrystalline anisotropy is easily related to $H_C$ in the isotropic limit but incoherent and inhomogeneous anisotropies are more difficult to quantify.  Magnetostatic energy is extrinsic, depending upon the shape of the magnet as well as the relative positions of particles in a powder measurement, and can also introduce anisotropy.  Collections of particles can reduce magnetostatic energy across a boundary without superexchange, while isolated particles must relax internally where superexchange competes with demagnetization; here we only model isolated particles.    Local anisotropies are present in NiCrPB near defects and at the surface (i.e. the aforementioned [Ni(NC)$_5$(H$_2$O)] species), depending upon the chemistry involved.  Indeed, CoFePB particles of similar sizes but different surface coordinations have strikingly different coercivities.\cite{p10}

\begin{figure*}[!]
	\includegraphics{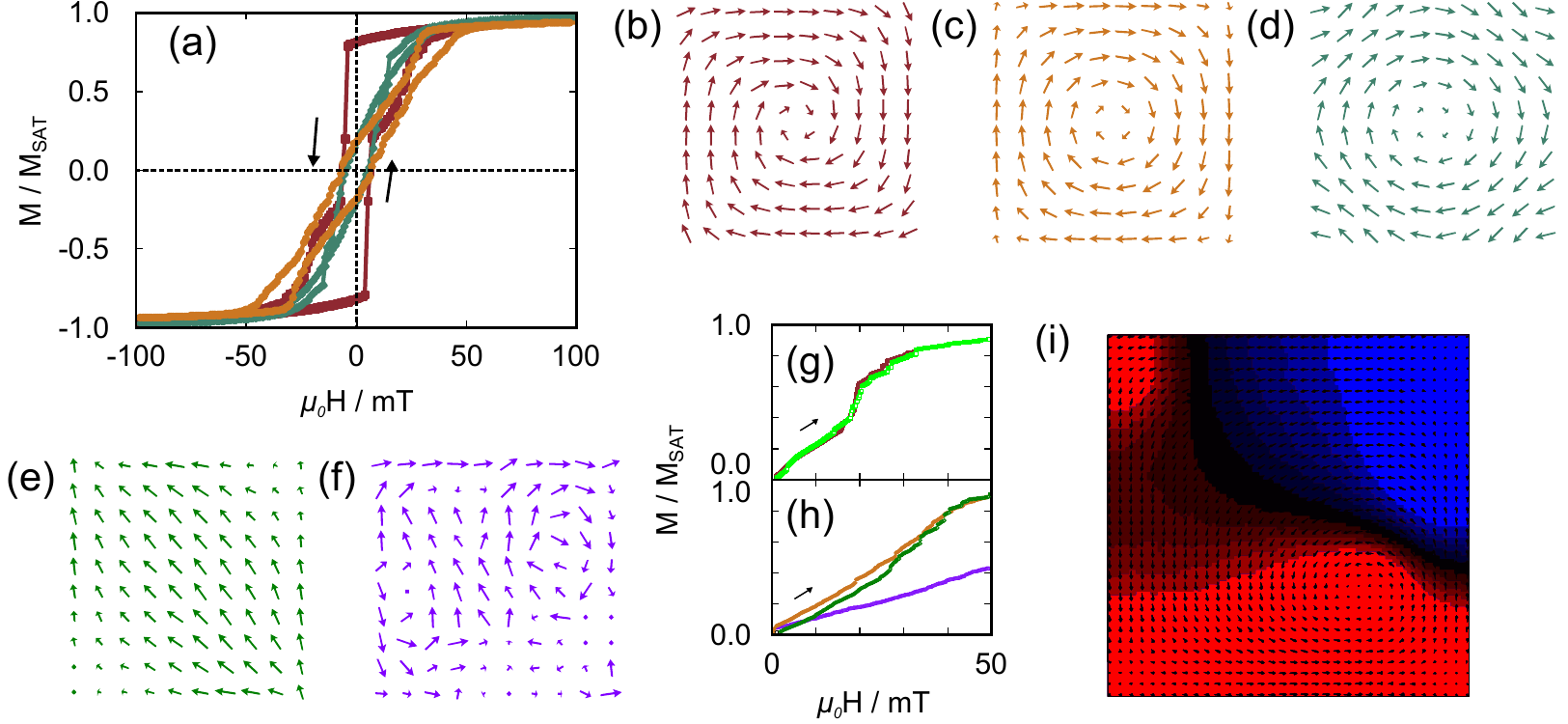}
	\caption{(color online) Micromagnetic simulations.  (a) Powder-averaged hysteresis loops are shown for Vol as burnt umber squares, SurfHard as ochre circles, and SurfEasy as viridian diamonds, with details described in the text.  Zero-field vortex states as a cut through the center of a particle are shown for (b) Vol, (c) SurfHard, and (d) SurfEasy, where states are nearly the same aside from the surface configuration.  (e) After a vortex is annihilated (or before it is created), the spins have a second reorientation process to undergo before saturation.  Here is a cut through the center of a particle in the x-z plane with an applied field of 33 mT applied along the (0.64 0.09 0.76) on SurfHardRand1 in the process of sweeping to zero after saturation.  The subtle incoherence to spin alignment due to the random anisotropy is present.  (f) For SurfHardRand5 at 20 mT along the (0 0 1) cut through the center of the particle in the x-y plane, local anisotropy dominates and the vortex is on the verge of being indiscernible.  (g) The initial powder-averaged magnetization of Vol, in burnt umber squares, shows a cross-over in magnetizing due to the vortex that can be shifted to higher field by including a single micromagnetic cell that increases pinning as in VolTron, in open bright green squares.  (h) Magnetizing becomes progressively harder with increased random anisotropy as shown for SurfHard as ochre circles, SurfHardRand1 as open green circles, and SurfHardRand5 as purple triangles.  (i) The SurfHardRand5 simulation with 1 nm$^3$ micromagnetic cells cut through the particle 3/4 along the y-axis, showing the x-z plane at zero field after returning from saturation along the (0.64 0.09 0.76).  Cell magnetizations have been down sampled three times to improve visibility and the coloring denotes magnetization magnitude out of the plane.}
	\label{fig:NiCr_figure7}
\end{figure*}

Therefore, we consider three models that capture the essence of the motivated coherent anisotropy distributions: (Vol) a constant cubic volume anisotropy, (SurfHard) uniaxial surface anisotropy having the hard axis normal to the surface by only applying anisotropy to surface micromagnetic cells, and (SurfEasy) uniaxial surface anisotropy having the easy axis normal to the surface by only applying anisotropy to surface micromagnetic cells.  In NiCrPB, for Ni:Cr~=~1:1, $T_C$~$\approx$~90~K and $\mu_0H_C$~$\approx$~7 mT  \cite{gmcvp92} and for Ni:Cr~=~1:0.75, $T_C$~$\approx$~70~K and $\mu_0H_C$~$\approx$~10~mT.\cite{p10}  To have a powder-averaged $\mu_0H_C$~$\approx$~7 mT, the anisotropy constants are $K_{1,Vol}$~=~7~$\times$~10$^3$~J/m$^3$, $K_{1,SurfEasy}$~=~1~$\times$~10$^4$~J/m$^3$, and $K_{1,SurfHard}$~=~-3~$\times$~10$^4$~J/m$^3$, with hysteresis loops shown in Fig.~\ref{fig:NiCr_figure7}(a).  Notably, the harsh inflection points in these simulated loops are not seen in most reported NiCrPB measurements, but are present in a diluted powder that better approximates isolated particles,\cite{cgb03} and finite temperature can also change magnetization.  From these values we estimate relevant anisotropies in NiCrPB are of the order 10$^4$ J/m$^3$, which corresponds to D~$\approx$~0.1~meV ($\approx$1~K) that is the same order of magnitude as the crude single-ion calculations presented in the previous section.  Micromagnetic parameters of NiCrPB are summarized in Table~\ref{table2}, and are quite different from typical parameters for bulk iron.\cite{s95}

The first micromagnetic result is that for all chosen NiCrPB anisotropy distributions, a magnetic vortex is stabilized in sufficiently low fields because of the demagnetizing energy.  In zero field, with an overall cubic symmetry vortices tend to be centered with boundary conditions determined by the surface anisotropy, Fig.'s~\ref{fig:NiCr_figure7}(b)-(d).  An applied field will first orient the vortex core and typically push it to one side as majority spins overtake minority spins.  The precipitous shape changes in the hysteresis loop are due to vortex formation and destruction.  The presence of such vortices (or other domain type) explains the low remanent magnetization ($M_R$) in NiCrPB, $M_R~\approx~0.1M_{SAT}$,\cite{p10} compared to magnets with uniaxial coherent rotation, $M_{R,uni}~=~0.5M_{SAT,uni}$, or cubic coherent rotation, $M_{R,cub}~\approx~0.8M_{SAT,cub}$.\cite{s95}

The second micromagnetic result is that anisotropy inhomogeneities can cause magnetization pinning of NiCrPB.  For the Vol particle, the anisotropy is essentially homogeneous and the introduction of only one micromagnetic cell with an additional uniaxial anisotropy of 10$^4$~J/m$^3$ (a simulation we called VolTron) can shift the de-pinning energy of the ground state vortex.  For SurfEasy and SurfHard, there already exists a region of strong inhomogeneity that causes pinning at the surface, such that a single impurity has no obvious effect.  Both types of surface anisotropy respond to random anisotropy in a similar way, so we only discuss SurfHard here.  A model system called SurfHardRand1 that introduces random axis anisotropy of random strength between 0 and 10$^4$~J/m$^3$ can reduce the low-field magnetization without affecting the high field magnetization or coercivity.  Increasing the strength of the random anisotropy to a maximum value of 5~$\times$~10$^4$~J/m$^3$ (SurfHardRand5) has a more profound effect as the anisotropy and superexchange energies become comparable; the coercive field is also increased ten times and saturation is pushed to higher fields.    However, SurfHardRand5 is more like a mosaic of weakly interacting uniaxial nanoparticles that undergo coherent rotation, as the micromagnetic cell size is 1000~nm$^3$.  After a vortex is annihilated, the spins must still continue to undergo rotation until saturation is reached, Fig.~\ref{fig:NiCr_figure7}(e).  In the SurfHardRand5, the chirality of the low field state is barely discernible due to the domination of random local anisotropy, Fig.~\ref{fig:NiCr_figure7}(f).  The sensitivity of the pinning field in Vol, to VolTron, is shown in Fig.~\ref{fig:NiCr_figure7}(g), and the effect of random anisotropy on SurfHardRand in the initial magnetizing region is shown in Fig.~\ref{fig:NiCr_figure7}(h).

To validate the simulations, we investigated size dependence of the micromagnetic cell.  Powder-averaged field sweeps are computationally expensive, so we checked the 10~nm cell results against 1~nm cells for zero field, 5~mT, 1~T, and a remnant field for SurfHard, SurfEasy, and Vol configurations.   These checks confirm the vortex state (with limited cell size dependence for models without random anisotropy) and pinning of magnetization to anisotropy inhomogeneities, although it is interesting how finer graining allows for more gradual relaxation of spins through short wavelengths that even smooths out comparatively down sampled distributions; for SurfHard5 with 1 nm$^3$ micromagnetic cells the anisotropy is still distributed randomly over 10~nm sided cubes, but the magnetization within those cubes can relax and disrupts the obvious coherent rotation behavior that was seen with the larger cells.  The remnant magnetization of SurfHard5 with 1~nm cells after a 1~T field was applied along the (0.64 0.09 0.76) direction with a cut through the particle shows the presence of one clear vortex with a core at an angle defined by the previously applied field and different major and minor spin populations, Fig.~\ref{fig:NiCr_figure7}(i).

\subsection{Other Considerations}
This idea of complicated magnetic ground states in CPs is not new, and pioneering work was done on CoFePB\cite{pmma00}\cite{pmma00a} and vanadium tetracyanoethylene (VTCNE)\cite{mzhej93}\cite{zma94}\cite{ppej01}, among other systems, that revealed the importance of random anisotropy to define glassy ground states in many CPs that can have weak superexchange and structural disorder.  Although we emphasize the importance of nanostructure, additional insight can be gained by examining the analytical theories.  Specifically, the low field magnetic susceptibility for a ferromagnet containing random uniaxial anisotropies was evaluated to be $\chi _{CSG}$~=~$1/2\beta _r (15/(4H_r/H_{ex}))^3$, where $\beta_r$ $\propto$ D$_r$, $H_{ex}~\propto~$A$ M_{SAT}/R_a$, $H_r = \beta _rM_{SAT}$, D$_r$ is a randomized D, and $R_a$ is the spatial scale over which anisotropy changes.\cite{csr86}  Therefore, as  random anisotropy in NiCrPB increases with pressure, the susceptibility from $\chi _{CSG}$ decreases, as is observed experimentally.  However, NiCrPB looks to be in the D$_r$~$<$~$J$ regime, which is different than CoFePB and solvent based VTCNE, which have D$_r$~$\approx$~$J$.  For D$_r$~$<$~$J$, coherent lab or anisotropy fields can easily anneal the glassy behavior.

While the largest relative change in magnetic response with pressure for NiCrPB is at low field, there is a more subtle change in high field that might be due to g-factor modification or large random anisotropy.  The effect of pressure induced structural distortions on spin-waves is less obvious, but presumably local hardening provides excitation gaps that would decrease the density of states at low temperatures.  From perturbation theory for Ni$^{2+}$ in a strong octahedral field, g$_{oct}$ = g$_{free}$ + 8$\lambda_{SO}^2$/$\Delta _{oct}$, and for axial distortions g$_{par}$ = g$_{free}$ + 8$\lambda_{SO}^2$/$\Delta _{0}$ and g$_{perp}$ = g$_{free}$ + 8$\lambda_{SO}^2$/$\Delta _{1}$, where g$_{free}$ is the free-ion g-factor, and $\Delta_0$ and $\Delta_1$ are the gaps to the $L_z=0$ and $L_z=\pm1$ excited states that are close in value to $\Delta_{oct}$.\cite{g61}  If [Ni(NC)$_6$] transforms to [Ni(NC)$_5$(CN)], the splitting between $t_{2g}$ and $e_g$ electrons can increase and reduce the g-factor.  Large random anisotropy could cause the high field spins to cant along a local axis and decrease the moment along the field.  While both explanations are plausible, the g-factor renormalization explanation is favored because any anisotropies that are strong enough to modify the high field magnetization seem to affect the coercive field and magnetic correlation length in ways inconsistent with experiment.

Better understanding of the magnetization in NiCrPB and how it responds to pressure can shed light on strain coupled heterostructures that exploit NiCrPB’s magnetoelasticity.  First, there is a definite extrinsic character to NiCrPB as demagnetizing fields play an important role in domain formation.  Indeed, even at high fields, demagnetizing effects were found to be important in NiCrPB.\cite{pgadgkhtm10}  On the other hand, the exchange length is essentially intrinsic and while the domain wall width and single-domain radius are extrinsic, the generic values in Table~\ref{table2} are useful to consider.  Then, it is not surprising that $\approx100$~nm layered heterostructures can show the largest magnetic response,\cite{pgfadkmd11} as that is close to the domain size.   So, domain wall formation in NiCrPB could be modified with the optically controlled strain, giving rise to the synergistic response.  Moreover, NiCrPB strain coupled heterostructures with $\leq$10~nm layers that show a seemingly opposite response to strain are actually in a different size regime than the $\approx100$~nm structures for both NiCrPB strain and magnetism.\cite{dlpgsbhsmct13}  Notably, nanoparticles of NiCrPB showed slightly different response to pressure than the “bulk” material, but a surfactant was used that might modify the surface anisotropy as well as the effective mulk modulus.\cite{zzkmz08}  Using our model of NiCrPB, each system must be approached individually to properly identify the magnetic ground state of the constituent particles even though the coordination polymer repeat unit might be identical.

So, this analysis of NiCrPB provides information that can be used by researchers engineering nanoscale magnets of CPs.  We invoke magnetic vortices and CN isomerism to explain the observables, but independent of these likely models, there is long correlation of magnetism in NiCrPB at low fields and pressure causes a striking change to the local magnetic ion environments.  While linkage isomerism of PBA compounds has been studied extensively with spectroscopic methods,\cite{vg05} which are sensitive for systems that have a concurrent spin transition, neutron diffraction directly measures differences in C and N site occupations.  To further support the model, it will be useful to check the NPD response of other systems, such as FeCrPB, that also have structural isomerization with pressure.  The NPD, and magnetic state, of NiCrPB is different than the previously studied CoFePB, which did not have magnetic scattering at low fields and is better classified as a coherent spin glass.\cite{pgkadcnttm12}  The common theme for NiCrPB and CoFePB is that a typical magnetic structure of one unit cell does not adequately describe the relevant observables.  We hope that these findings provide insight to CP researchers not only using NiCrPB, but also for analogous systems that are in the same magnetic parameter regime and can support complicated nanomagnetism.

\section{Conclusions}
We have presented a model for the magnetizing process in NiCrPB that invokes magnetic vortex domains, and the application of strain introduces inhomogeneous anisotropy via CN-isomerism of $\approx$20\% of the sites that increases domain pinning energies.  The general picture of magnetization in CPs is becoming clearer due to NPD, and the complicated magnetism within NiCrPB underscores the need to congruently consider magnetic terms (ground and excited), local structure, and nanostructure when interpreting these systems.  Indeed, precedented analysis algorithms, such as the magnetic unit cell, may not be adequate to describe the complex behavior of CPs.  For the subclass of PBA CPs, this work further stresses the need to approach each system independently, even though analogies across analogues is often useful.

\begin{acknowledgments}
DMP would like to thank M. W. Meisel for critical reading of the manuscript, P. Kienzle for advice on code optimization, B. B. Maranville for advice on micromagnetic simulations, C. Majkrzak and K. Krycka for help setting up polarization analysis, B. B. Maranville and C. Majkrzak for discussions about the coherence length of the neutron, W. Ratcliff for use of BT-4, and P. M. Gehring for allowing use of a day of his neutron time and general encouragement.  SEC was funded by the National Institute of Standards and Technology Summer Undergraduate Research Fellowship.
\end{acknowledgments}


\bibliographystyle{apsrev4-1}
\bibliography{nicr}

\begin{thebibliography}{46}
\bibitem [{NIS()}]{NISTdisclaimer}%
  \BibitemOpen
  \href@noop {} {}\bibinfo {note} {Certain commercial equipment, instruments,
  or materials are identified in this paper to foster understanding. Such
  identification does not imply recommendation or endorsement by the National
  Institute of Standards and Technology, nor does it imply that the materials
  or equipment identified are necessarily the best available for the
  purpose.}\BibitemShut {Stop}%
\end{thebibliography}

\end{document}